\documentclass[twocolumn,
superscriptaddress,
prl,
showpacs,
preprintnumbers,
amsmath,amssymb]{revtex4}

\usepackage{graphicx,epsf,epsfig,ulem}
\usepackage{epsfig}
\usepackage{bm}
\usepackage{subfigure}
\usepackage{color}

\begin{document}

\title{Lifetime enhanced transport in silicon due to spin and valley blockade}

\author{G. P. Lansbergen*}
\affiliation{Kavli Institute of Nanoscience, Delft University of Technology, Lorentzweg 1, 2628 CJ Delft, The Netherlands}

\author{R. Rahman}
\affiliation{Advanced Device Technologies, Sandia National Laboratories, Albuquerque, NM 87185, USA}

\author{J. Verduijn}
\affiliation{Kavli Institute of Nanoscience, Delft University of Technology, Lorentzweg 1, 2628 CJ Delft, The Netherlands}
\affiliation{Centre for Quantum Computation and Communication Technology, School of Physics, University of New South Wales, Sydney, New South Wales 2052, Australia}

\author{G. C. Tettamanzi}
\affiliation{Kavli Institute of Nanoscience, Delft University of Technology, Lorentzweg 1, 2628 CJ Delft, The Netherlands}
\affiliation{Centre for Quantum Computation and Communication Technology, School of Physics, University of New South Wales, Sydney, New South Wales 2052, Australia}

\author{N. Collaert}
\affiliation{Inter-University Microelectronics Center (IMEC), Kapeldreef 75, 3001 Leuven, Belgium}

\author{S. Biesemans}
\affiliation{Inter-University Microelectronics Center (IMEC), Kapeldreef 75, 3001 Leuven, Belgium}

\author{G. Klimeck}
\affiliation{Network for Computational Nanotechnology, Purdue University, West Lafayette, IN 47907, USA}

\author{L. C. L. Hollenberg}
\affiliation{Centre for Quantum Computation and Communication Technology, School of Physics, University of Melbourne, VIC 3010, Australia}

\author{S. Rogge}
\affiliation{Kavli Institute of Nanoscience, Delft University of Technology, Lorentzweg 1, 2628 CJ Delft, The Netherlands}
\affiliation{Centre for Quantum Computation and Communication Technology, School of Physics, University of New South Wales, Sydney, New South Wales 2052, Australia}

\date{\today}

\begin{abstract}
We report the observation of Lifetime Enhanced Transport (LET) based on perpendicular valleys in silicon by transport spectroscopy measurements of a two-electron system in a silicon transistor. The LET is manifested as a peculiar current step in the stability diagram due to a forbidden transition between an excited state and any of the lower energy states due perpendicular valley (and spin) configurations, offering an additional current path. By employing a detailed temperature dependence study in combination with a rate equation model, we estimate the lifetime of this particular state to exceed 48 ns. The two-electron spin-valley configurations of all relevant confined quantum states in our device were obtained by a large-scale atomistic tight-binding simulation. The LET acts as a signature of the complicated valley physics in silicon; a feature that becomes increasingly important in silicon quantum devices.    

\end{abstract}


\maketitle 

Silicon, the most important material in the electronics industry till date, is starting to make its mark in nanoscale quantum electronics. While silicon nanowires are showing promise in thermoelectric applications \cite{Heath}, silicon based qubits \cite{Friesen1, Kane1, Vrijen, Hollenberg1, Hill} have been widely investigated worldwide due to their long spin coherence times \cite{Tryshkin1} and compatibility with the present-day infrastructure of the semiconductor industry. In recent times, both top-down  and bottom-up approaches have been successful in building single to few electron silicon devices \cite{Koenraad, Eriksson, Simmons2}. 
The demonstration of a single-shot readout of an electron spin \cite{Andrea} marks a crucial step towards building a silicon qubit. Unlike most materials used for qubits, silicon posses a complex momentum space structure, primarily arising due to the six degenerate conduction band (CB) valleys away from the Brillouin Zone center. The CB valley degeneracy provides an extra degree of freedom to the quantum confined orbital states in silicon, and gives rise to a variety of novel phenomena that strongly affect the electronic structure, transport, and relaxation mechanisms \cite{Eto, Sarma2, Kondo, Fujiwara, Sarma3}. It is therefore important to understand and quantify valley physics and its effect on electronic states. 

In this letter, we report observations of a novel valley induced phenomena in electronic transport in silicon. Through transport spectroscopy measurements of donor states in FinFETs, we show that under certain conditions relaxation of excited states into lower manifolds is suppressed due to a combination of both spin and valley blockade. This enhanced lifetime results in an additional transport path through the excited state, and appears as a current step in the stability diagram. The phenomena dubbed as `lifetime enhanced transport' (LET) was first observed in a silicon double quantum dot \cite{Eriksson} due to a blocked relaxation of a spin triplet into a ground state spin singlet, arising from the long spin relaxation times in silicon. In this experiment, LET enables us to identify a blocked transition between states that have different valley symmetries. We confirm this observation (i) by extracting the tunnel rates in and out of the donor states through a temperature dependent measurement and analysis, and (ii) by 
computing the low-energy two-electron spectrum of the system from a multimillion atom tight-binding method to compare and identify the measured excited manifolds.  

\begin{figure}[htbp]
\center\epsfxsize=2.4in\epsfbox{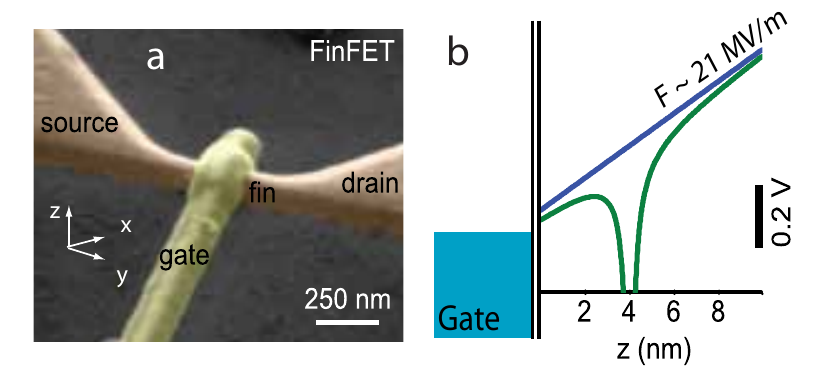}
\caption{a) Scanning electron micrograph (SEM) image of the FinFET device used in the experiment. b) A 1D Schematic of the donor potential under an applied electric field.}
\vspace{-0.5cm}
\label{fi1}
\end{figure}

The devices used in this study are Si FinFET transistors, consisting of a Si nanowire with a gate covering three faces of the body (Fig.1a). We have measured the low-temperature electronic transport of devices where the sub-threshold characteristics are dominated by a single As donor atom in the channel. The characteristic fingerprint of a single dopant consist of a pair of resonances associated with the one electron neutral donor ($D^0$) and two-electron negatively charged donor ($D^-$) states, a $D^0$ binding energy of $\sim$ 50 meV, a $D^-$ charging energy between 30-35 meV, and a spin odd-even effect \cite{Sellier}. 
The high electric field transforms the confining potential to a (hybridized) mix between the donor's Coulomb potential and a triangular well at the interface, as depicted in Fig. 1b) \cite{Calderon, Rahman}. 
This system is essentially a gated donor where the donor-bound electrons are partly pulled toward the Si/SiO$_2$ interface. LET is observed in the transport through the two-electron $D^-$ spectrum of the donor, as we will show below. 

In a previous work \cite{Rogge1}, we measured the Stark shifted $D^0$ excited state spectrum of an As donor in a number of devices, and through atomistic device modeling we inferred the precise locations of the donors and the fields they experienced. Although the charging energy (CE) of a bulk As donor is about 52 meV (corresponding to a binding energy of 2 mev for $D^-$), the CEs in these devices are reduced to 30-35 meV due to applied fields and capacitive coupling to the gates \cite{Sellier}. As a result, excited $D^-$ manifolds form below the conduction band, and are visible in the stability diagram. 

\begin{figure}[htbp]
\center\epsfxsize=3.2in\epsfbox{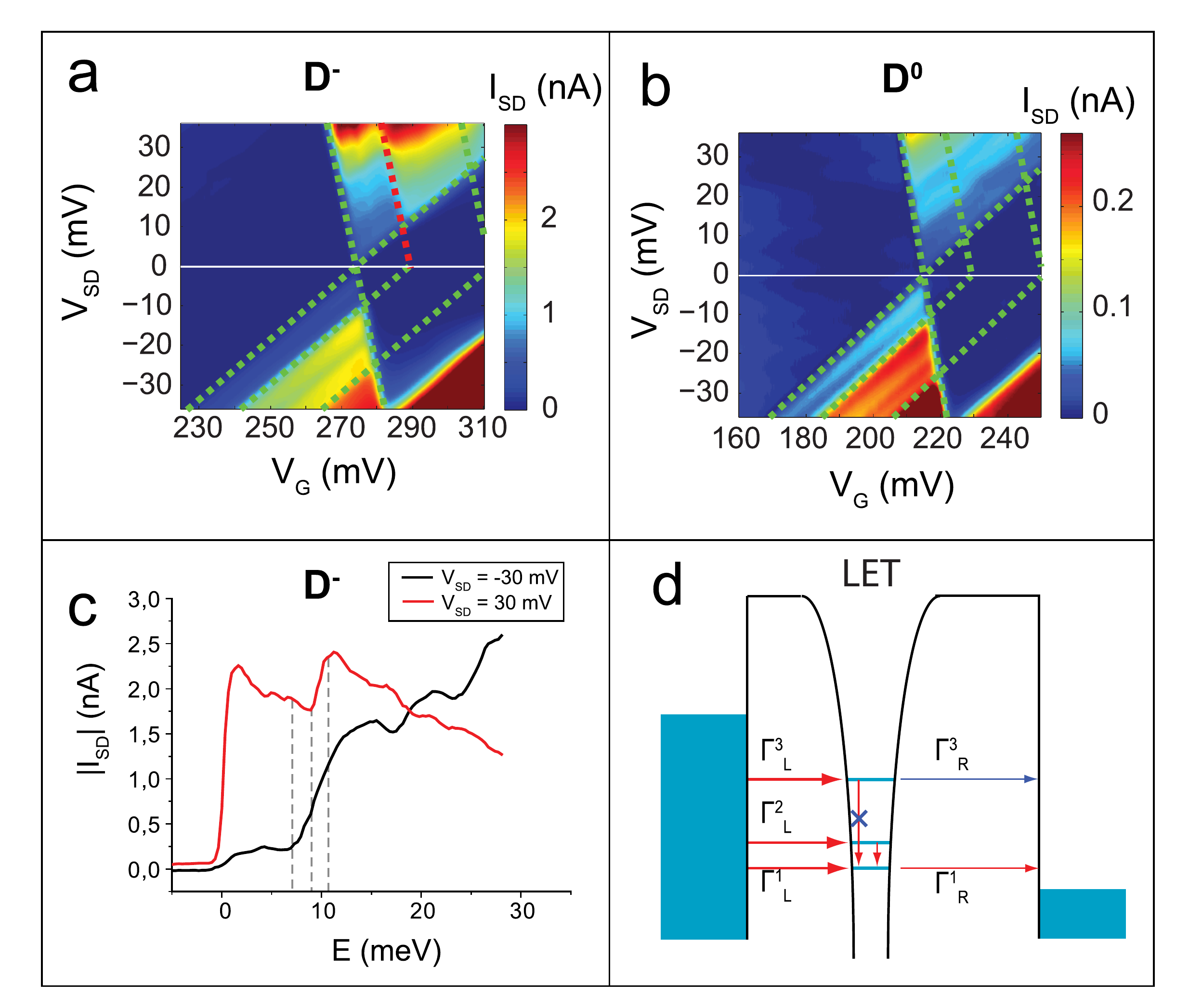}
\caption{Stability diagrams of the a) $D^-$ and b) $D^0$ charge state resonance. The green dashed lines indicate current steps associated with an excited state entering the bias window. The highly asymmetric coupling to the leads and the sequential nature of the transport quench the current steps in the V$_{\rm SD} > 0$ conducting region. The current step in the V$_{\rm SD} > 0$ indicated by the red dashed line in a) must have a life-time ($\tau) >$ 48\,ns to show up significantly in the DC transport. c) Source/drain current through the $D^-$ versus energy $E$ (with respect to the ground state) for $V_G$ = -30\.meV and $V_G$ = +30\.meV, showing the additional current step around 11 meV due to LET. d) Schematic of the transport processes through the donor atom. Sequential processes are characterized by relaxation to the ground state before tunneling out of the atom, indicated by the transport channel ${\Gamma_R}^1$. For LET processes, such as the one marked by the red dashed in line in a), the additional transport path ${\Gamma_R}^3$ comes into play. }
\vspace{-0.5cm}
\label{fi3}
\end{figure}

Fig. \ref{fi3}a and \ref{fi3}b show the stability diagram around the $D^-$ and $D^0$ charge state respectively. We observe several steps in the conducting region which can be associated with the excited state spectrum \cite{Rogge1, Klimeck1}, indicated by the green dashed lines. These steps are clearly present for V$_{\rm SD} < 0$ but are absent for V$_{\rm SD} > 0$ (except for one current step indicated by the red dashed line in Fig. \ref{fi3}a, an indication of strong {\it asymmetric coupling} with respect to the source and drain contacts . The peculiar current step in Fig. \ref{fi3}a marked by the red dashed line for V$_{\rm SD} > 0$ is the feature we will focus on. The only way it can occur is if the excited state has an unusually long life-time so that  it does not relax to the ground state during transport (Fig. \ref{fi3}d)  \cite{Klimeck1}. 

The position of the steps in the conducting regions of the Coulomb diamond reflect the energy of the excited states. The V$_{SD}<0$ trace of Fig. \ref{fi3}c 
shows that there is an excited state between 7.5\,meV and 15\,meV and another one at around 20\,meV. However, such steps do not necessarily have to originate from a single state, a set of quantum states can be too close in energy space to be distinguished separately. Actually, the second step in the conducting region is very broad, starting at 7.5\,meV and reaching its maximum height at around 15\,meV (Fig.\,2c), which points to an ensemble of states entering the bias window. This is confirmed by the trace at V$_{\rm SD} > 0$ where the single step is much narrower than at the other polarity and must belong to one of the states of this larger ensemble. As we will show by an analysis of the temperature dependence, the first step associated with the ground state also originates from multiple states; there is at least one first excited state at $\sim$1\,meV. These experimental values correspond quite well with the energy spectrum of the two-electron donor calculated from a tight-binding based Hartree Fock approach, as shown later. 

\begin{figure}[htbp]
\center\epsfxsize=3in\epsfbox{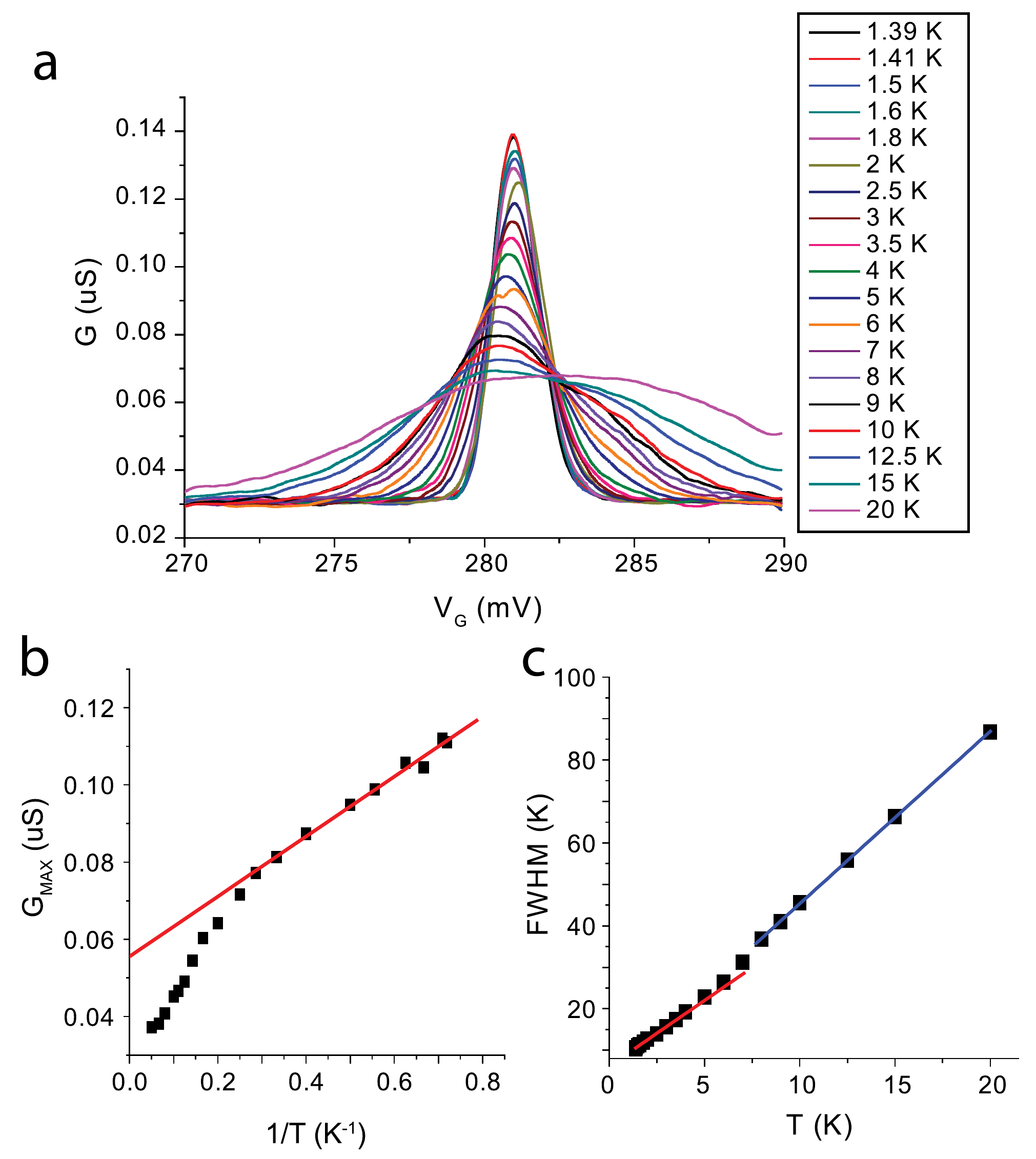}
\caption{Temperature dependence of the ground state resonance. a) Peak shape at various temperatures. b) Peak height (G$_{\rm MAX}$) and c) FWHM of the resonance as obtained by a Gaussian fit. The change in behavior of b) and c) at the transition temperature T$_C=5$ K shows the crossover from single to multi-level transport regime \cite{Foxman}.  
}
\vspace{-0.5cm}
\label{fi2}
\end{figure}



A temperature dependent analysis of the ground state (GS) conductance peak (G$_{\rm MAX}$) helps us to verify that the transport is in the sequential regime, and enables us to extract the tunneling rates $\Gamma_L$ and $\Gamma_R$ between donor and the left and right leads (Fig. \ref{fi3}d) \cite{Foxman}.
Fig. \ref{fi2}a shows the GS resonance at various temperatures. The resonance broadens with increasing temperature due to the temperature-dependent Fermi-Dirac distribution in the leads. The resonances were fitted with Gaussian functions to extract the the peak height (G$_{\rm{MAX}}$) and the Full-Width-Half-Maximum (FWHM), as shown in Fig.\,\ref{fi2}b and Fig. \ref{fi2}c respectively. Both figures show a transition in the transport regime roughly around 5\,K. Below this transition temperature T$_C$, G$_{\rm{MAX}}$ increases linearly with $1/{\rm T}$, and the FWHM increases linearly with T. This specific temperature dependence fully characterizes the transport as being {\it sequential} and dominated by a single level (i.e. the GS). The slope of the FWHM as a function of T is equal to 3.2$\pm$\,0.2\,$k_{B}$, very comparable to the theoretical value of 3.5\,$k_{B}$ \cite{Foxman}. The small discrepancy is attributed to a finite lifetime broadening in our system. Above the transition temperature, the slope of the FWHM as a function of T is equal to 4.16\,$\pm$\,0.02\,$k_{B}$, comparable to the theoretical value of 4.35\,$k_{B}$ \cite{Foxman} for transport dominated by multiple levels. The small discrepancy is attributed to the fact that the theoretical value holds for equidistant level spacing, which is not the case in our system. 
The transition temperature corresponds to the first excited state starting to contribute to the transport, which yields a splitting between the GS and first excited state of $k_{B}{\rm T}_C \sim$ 1\,meV. 

We can also obtain the coupling of the GS to the source and drain leads from the temperature dependence of (G$_{\rm{MAX}}$ below T$_C$, which is given by \cite{Klimeck1, Beenakker}) \begin{equation}
  G_{\rm MAX}=\frac{e^2}{4k_{B}T}\frac{\Gamma_L \Gamma_R}{\Gamma_L + \Gamma_R} \label{vgl1}
\end{equation}
By fitting G$_{\rm{MAX}}$ below T$_C$ to eq.\,\ref{vgl1} we find $\frac{\Gamma_L \Gamma_R}{\Gamma_L + \Gamma_R} = 4.2 \times 10^{13}$. Next, we have estimated $\Gamma (= \Gamma_L + \Gamma_R)$ from transport measurements at 300\,mK, where $\Gamma > k_{B}{\rm T}$ and the transport is thus in the {\it coherent} regime. Here the FWHM of the resonance is mainly lifetime-broadened, i.e. FWHM = $\Gamma$ \cite{Buttiker}, and we found $\Gamma$ = 173 $\mu$eV. Combining the latter two pieces of information yields $\Gamma_L$ = 173 $\mu$eV and $\Gamma_R$ = 86 neV. This shows that the donor state in this device is coupled asymmetrically with the source and drain, with a ratio $\Gamma_L/\Gamma_R\approx 2000$.

A rate equation model now verifies the observed voltage asymmetry and identifies the peculiar current step of Fig. \ref{fi3}a as arising from LET. 
In the sequential transport regime, the current through the donor atom is given by \cite{ChemPhysChem} 
 \begin{equation}
 I = e \frac{\left( \Gamma_{\rm in}^1 + \Gamma_{\rm in}^2 + ... + \Gamma_{\rm in}^n \right) \Gamma_{\rm out}^1}{\Gamma_{\rm in}^1 + \Gamma_{\rm in}^2 + ... + \Gamma_{\rm in}^n + \Gamma_{\rm out}^1} \label{vgl2}
\end{equation} 
where the subscript denotes the tunneling rate in or out of the donor, the superscript indicates the energy level (1 represents the GS) and $n$ indicates the number of states in the bias window defined by the source-drain bias. Thus the electron can enter the donor through any number of states in the bias window, but leave only through the GS, as it eventually relaxes to the GS before exiting. Since the donor in our sample is very asymmetrically coupled to the leads, $\Gamma_{\rm in} >> \Gamma_{\rm out}$ for V$_{\rm SD} >$ 0 while $\Gamma_{\rm in} << \Gamma_{\rm out}$ for V$_{\rm SD} <$ 0. In this asymmetric limit Eq.\,\ref{vgl2} reduces to 
\begin{eqnarray}
  I =  \lbrace & e \left( \Gamma_{\rm in}^1 + \Gamma_{\rm in}^2 + ... + \Gamma_{\rm in}^n \right) & V_{\rm SD} < 0  \\
                 & e \Gamma_{\rm out}^1 & V_{\rm SD} > 0 \rbrace 
 \label{vgl3}
 \end{eqnarray}
This shows that if $V_{\rm SD}<0$, the current steps increase as more states enter the bias window. On the other hand, if $V_{\rm SD}>0$, the current becomes saturated and no current steps are expected. Actually, with our degree of asymmetry the maximum magnitude of any current step ($n > 1$) in $V_{\rm SD} > 0$ equals $0.5\,\%$ of the step at $n=1$, regardless of $\Gamma^n$. The latter follows directly from Eq.\,\ref{vgl2} in combination with the asymmetry $\Gamma_L / \Gamma_R \approx$ 2000 we found from the temperature dependent data (Fig. \ref{fi2}). Thus the current step due to LET for $V_{\rm SD}>0$, shown in Fig. \ref{fi3}a as the red line marks a deviation from the sequential transport regime, as an extra current path opens up through an excited state. By the magnitude of the current steps at $V_{\rm SD} < 0$ we can (conservatively) estimate $\Gamma_{R}^3 \sim \Gamma_{R}^1$, thus the life-time $\tau >$ 48\,ns.



\begin{figure}[htbp]
\center\epsfxsize=3.2in\epsfbox{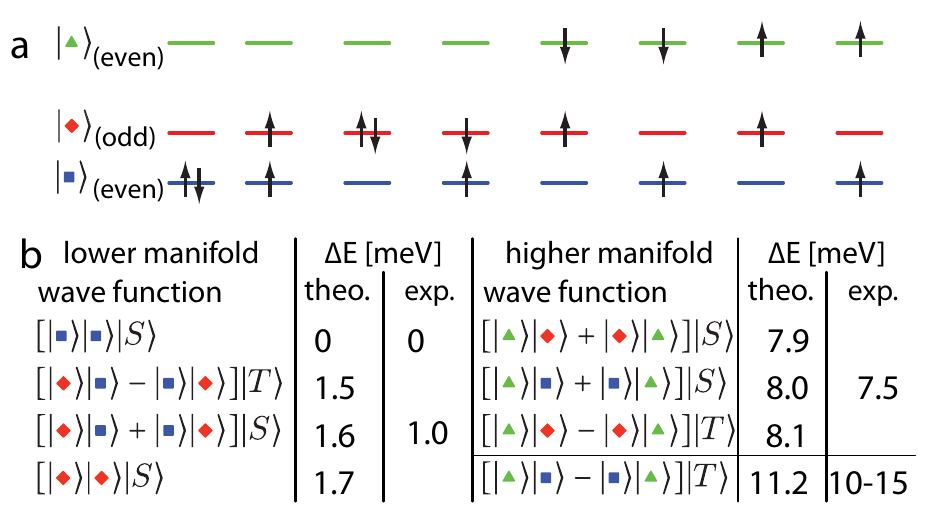}
\caption{Calculated spin-valley configuration of the $D^-$ level spectrum. 
a) Result of the Hartree-Fock tight-binding calculations and comparison with the experimental data. The energies are labeled relative to the 2e ground state energy. b) The spin( up/down) and valley (odd/even) configuration of 1e states showing the generated 2e configurations. From $D^0$ transport spectroscopy of the system, the donor depth and field have been determined to be 4.7 nm and 23 MV/m respectively \cite{Rogge1}.}
\vspace{-0.5cm}
\label{fi4}
\end{figure}

To identify the spin and valley configurations of the $D^-$ excited states, we performed a two-electron (2e) Hartree Fock calculation using good one-electron (1e) wavefunctions obtained from atomistic tight-binding simulations \cite{Klimeck2} of the device. 
In the bulk, a donor ground state has equal contributions from all six CB valleys, however, the donor in this sample is under a strong z-directed field, which makes the $k_z$ valley contributions dominant in the lowest few 1e wavefunctions, as we also verified by computing the Fourier Transform of the wavefunctions. Since there is strong valley-orbit interaction arising from the sharp interfacial potential and the central-cell donor potential, the two $k_z$ valley states are in symmetric (even) and anti-symmetric (odd) superposition \cite{Rahman}. The lowest three orbital states shown in Fig. 4a as $| \Box \rangle$, $| \Diamond \rangle$, and $| \triangle \rangle$  correspond to even, odd, and even valley symmetries respectively. 15 different 2e configurations (i.e. Slater Determinants) can be formed from these lowest three orbital (six spin) states, their spin symmetries being either singlet or triplet. Some of the non-degenerate 2e configurations are shown in Fig. 4a, marking the occupation of the electrons in the spin-valley configurations. 


 
In Fig. 4b, we compare the 2e spectrum from theory and measurement. Consistent with the measured data of Fig. 2b and 2c, we obtain manifolds of states around 1 meV, 8 meV, and 11 meV, each manifold containing both singlet and triplet states. The lower manifold of states ($<$ 2\,meV) does not contain a state with a triplet spin- and same (even-even or odd-odd) valley -combination, as a triplet spin is not possible within the same valley. The first triplet spin- same valley -configuration state in the spectrum is theoretically at 11.2\,meV ($| \triangle \rangle | \Box \rangle + | \Box \rangle | \triangle \rangle$$|T\rangle$ in Fig.\,4b). This particular state can not relax back to any of the lower states in the spectrum, before either flipping spin (triplet to singlet) or flipping valley configuration (even-even to odd-even). The $| \triangle \rangle | \Box \rangle + | \Box \rangle | \triangle \rangle$$|T\rangle$ states are very close in energy to the experimental lifetime enhanced state at 10\,meV. Thus the observed LET is due to a state with triplet spin- and same valley -configuration, present in the higher manifold of states. The state is thus life-time enhanced by both spin- and valley -blockade.

In the bulk, the spin relaxation time of donor-bound electrons if of the order of milli-seconds, 
however, the vicinity of the Si/SiO$_2$ interface is expected to reduce that time-scale. We can put a (conservative) lower bound on the singlet-triplet relaxation in our $D^-$ system by comparing it to the relaxation time T$_1$ of 400 nm quantum dots defined in a two-dimensional Si/SiO$_2$ electron system measured by spin echo \cite{Lyon}, which was about 460\,ns. 
In a silicon 2D confined geometry, valley-lifetimes in the order of $\mu$s are to be expected \cite{Sarma2}. This is all consistent with the lower bound of 48 ns valley-lifetime estimated here. 

In this letter, we reported the observation of LET in silicon arising both due to spin and valley selection rules. We showed that in the sequential transport regime, LET can be manifested as peculiar current steps in the stability diagram as excited states fail to relax to lower states of different valley symmetries, resulting in additional transport paths. The unusually long-lived states of different valley symmetries allude to the fact that the valley index can form a two-level system and thus can act as a pseudo-spin quantum number. It is interesting to note that this pseudo-spin has recently been proposed as the basis in a {\it valley}-based quantum computation scheme \cite{Sarma3}. The LET thus serves as a signature of of the complex momentum space of silicon, and highlights the importance of understanding valley physics for silicon quantum electronics.

\begin{acknowledgments}
This work was funded by the Australian Research Council Centre of Excellence for Quantum Computation and Communication
Technology (Project No. CE110001027) 
and the Army Research Office (Contract No. W911NF-04-1-0290). NEMO 3D was developed at JPL, Caltech under a contract with NASA. Sandia is a multiprogram laboratory operated by Sandia Corporation, a Lockheed Martin Company, for the United States Department of Energy's National Nuclear Security Administration under Contract No. DE-AC04-94AL85000. NCN/nanoHUB.org resources were used. 
\end{acknowledgments}

Electronic address: gabriel.lansbergen@lab.ntt.co.jp

\end{document}